\documentclass[conference]{IEEEtran}
\IEEEoverridecommandlockouts
\usepackage{cite}
\usepackage{amsmath,amssymb,amsfonts}
\usepackage{algorithmic}
\usepackage{graphicx}
\usepackage{textcomp}
\usepackage{float}
\usepackage{tabularx}
\usepackage{array}
\usepackage{multirow, makecell}
\usepackage{xcolor}
\usepackage{caption}

\def\BibTeX{{\rm B\kern-.05em{\sc i\kern-.025em b}\kern-.08em
    T\kern-.1667em\lower.7ex\hbox{E}\kern-.125emX}}
\begin{document}

\title{Deep Learning for Musculoskeletal Image Analysis}

\author{\IEEEauthorblockN{Ismail Irmakci \IEEEauthorrefmark{1}, Syed Muhammad Anwar \IEEEauthorrefmark{2}, Drew A. Torigian \IEEEauthorrefmark{3}, Ulas Bagci \IEEEauthorrefmark{2}}


\IEEEauthorblockA{\IEEEauthorrefmark{1} \textit{Electrical and Electronics Engineering} \\
\textit{Ege University}\\
Izmir, Turkey. \\
ismail.irmakci@ege.edu.tr
}
\IEEEauthorblockA{ \IEEEauthorrefmark{2} \textit{Center for Research in Computer Vision} \\
\textit{University of Central Florida}\\
Orlando, Florida, USA. \\
s.anwar@knights.ucf.edu, bagci@ucf.edu \\
}
\IEEEauthorblockA{ \IEEEauthorrefmark{3} \textit{Department of Radiology} \\
\textit{University of Pennsylvania}\\
Philadelphia, Pennsylvania, USA. \\
drew.torigian@pennmedicine.upenn.edu
}

}

\maketitle


\begin{abstract}
The diagnosis, prognosis, and treatment of patients with musculoskeletal (MSK) disorders require radiology imaging (using computed tomography, magnetic resonance imaging (MRI), and ultrasound) and their precise analysis by expert radiologists. Radiology scans can also help assessment of metabolic health, aging, and diabetes. This study presents how machine learning, specifically deep learning methods, can be used for rapid and accurate image analysis of MRI scans, an unmet clinical need in MSK radiology. As a challenging example, we focus on automatic analysis of knee images from MRI scans and study machine learning classification of various abnormalities including meniscus and anterior cruciate ligament tears. 
Using widely used convolutional neural network (CNN) based architectures, 
we comparatively evaluated the knee abnormality classification performances of different neural network architectures under limited imaging data regime and compared single and multi-view imaging when classifying the abnormalities. Promising results indicated the potential use of multi-view deep learning based classification of MSK abnormalities in routine clinical assessment.
\end{abstract}


\section{Introduction}
Musculoskeletal (MSK) imaging is a sub-specialty in clinical radiology dealing with diagnosis of diseases related to bones and soft tissues. There are different imaging modalities used in MSK radiology including computed tomography (CT), magnetic resonance imaging (MRI) and ultrasound. In general, MRI is the preferred modality in most cases when presented with multiple imaging options, owing to its strong capability in providing an excellent soft tissue contrast and its radiation-free nature \cite{nacey2017magnetic}. In diagnostic evaluations of scans, radiologists image joints in different body regions such as knee, hip, and shoulders, tissues such as bone and muscle, assess peripheral nervous system, and perform whole body imaging for detecting disorders in metabolism, aging, and diabetes. The spectrum of clinical diagnosis using these radiology images is wide and continuously expanding with improved imaging sequences and associated computational methods \cite{gyftopoulos2019artificial}. Among various MSK radiology applications, knee MRI is found to be dominating, since the potential of accurately detecting abnormalities in the knee region is found to be higher \cite{bien2018deep}. There are multiple abnormalities that can be detected by using knee MRI including meniscal and cruciate pathology and cartilage \cite{raj2018automatic}. The number of scans as well as the amount of information present within a knee MR are the driving forces behind the need for developing automated analysis techniques in such clinical applications.  

Machine learning, specifically deep learning models, have been successful in a wide range of image analysis tasks. Medical image computing has been benefiting from these advances and resulting in some exciting avenues of research \cite{anwar2018medical}, \cite{anwar2019survey}. One of the advantages of using these deep learning algorithms is their ability to learn an effective data representation without the need for pre-identifying the appropriate features in a hand crafted manner. Although there has been an expansion in the number of applications and methods proposed for automated radiological image analysis, there are certain challenges that have started to surface \cite{anwar2019artificial}, \cite{esteva2019guide}. A pressing challenge in the field of medical imaging is the lack of large amount of imaging data and their precise annotations for effectively training deep learning models. Despite these considerations, deep learning powered methods still continue to revolutionize medical imaging and are expected to significantly affect the field of radiology \cite{mcbee2018deep}.

\begin{figure*}[!ht]
\centerline{\includegraphics[width = 180mm]{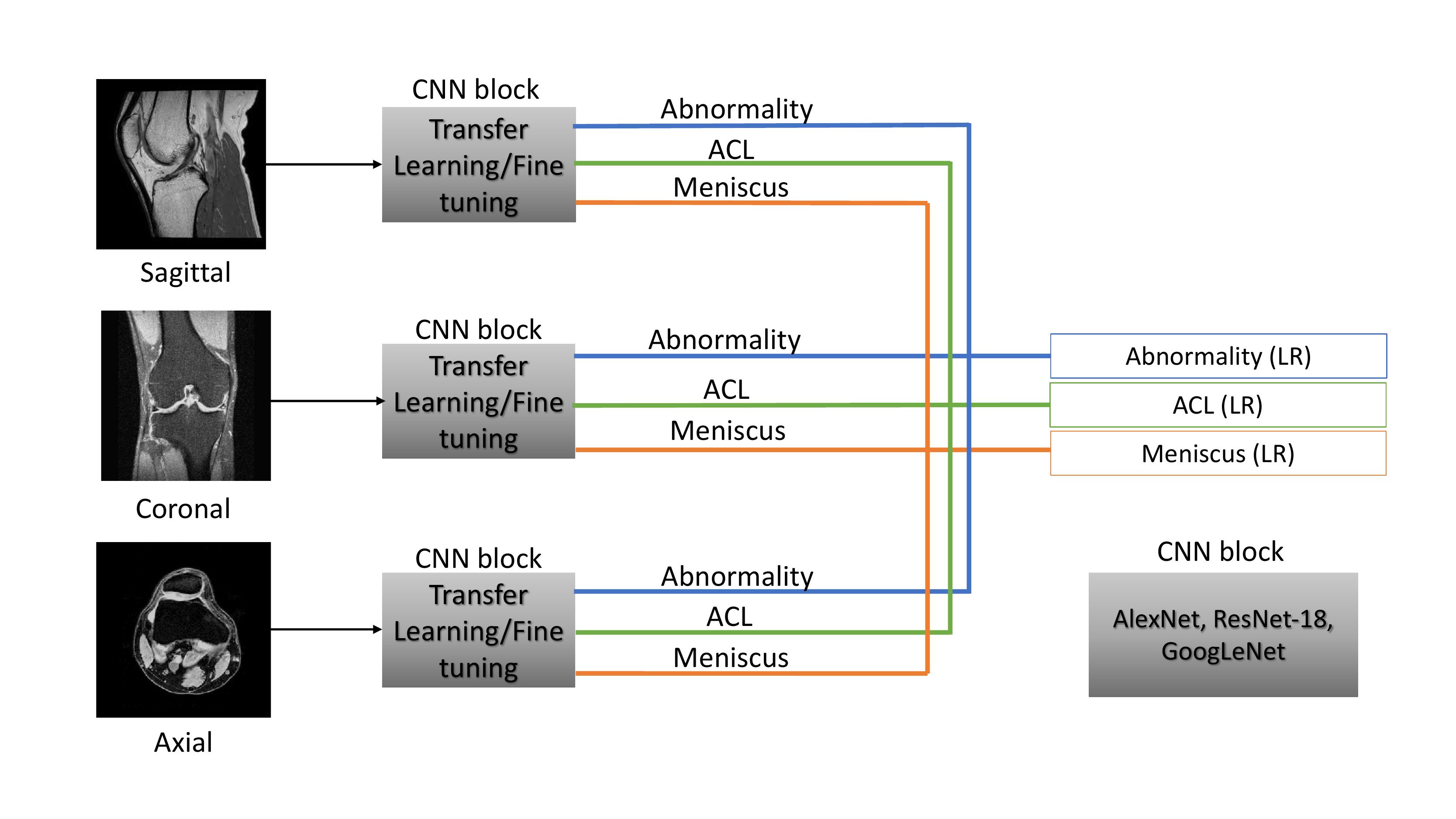}}
\caption{A block diagram of the proposed methodology is shown. For each view (plane), there is a separate CNN fine tuned for the classification problem. Different architectures were used in the same problem for comparison (AlexNet, ResNet, GoogLeNet). }
\label{fig}
\end{figure*}

There are various studies reported in literature that have used MSK radiology scans and pre-deep learning approaches to predict clinical conditions  \cite{chalian2016second}. Among these predicting muscle strength has been one of the major areas of study. The effect of facioscapulohumeral dystrophy (FSHD) on muscle strength was evaluated using MRI scans and musculoskeletal modeling for patients and healthy volunteers \cite{marra2018specific}. It was concluded that fatty infiltration and atrophy could be the factors that cause muscle weakness in patients with FSHD. In this particular direction, quantification of fat within the muscles and other body regions is important. There are methods developed using machine learning techniques for the segmentation of such tissues in the thigh region \cite{irmakci2018novel}. More advanced techniques (such as those based on deep learning) could further benefit the clinical progress in this field. In musculoskeltal knee MRI the image could contain a huge amount of information, whereas the most commonly used clinical sequences (such as 2D fast spin echo) are prone to miss information due to high slice thickness. Super resolution MRI is a way to generate thin slices form these thick slices, having significantly better resolution. In a recent study, it has been shown that deep learning helped in generating such super resolution MRIs \cite{chaudhari2018super}. The proposed 'DeepResolve' method gave better qualitative and quantitative results in improving image resolution. Convolutional neural networks and 3D deformable models were used to segment cartilage and bone in the knee MRI \cite{liu2018deep}. It was shown that fully automated segmentation of musculoskeltal MRI was possible using such methods for rapid and accurate cartilage segmentation in the knee region.

Machine learning based prediction approaches could help in solving existing problems in musculoskeletal radiology and lead towards new and exciting areas of research. The performance of machine learning based algorithms can be improved by using more specialized deep learning models which are fine tuned to a particular application area. It also needs to be explored how deeper convolution neural networks would adapt to this particular area of analyzing knee MRI. In this regard, to facilitate development of deep learning models for detecting abnormalities from knee MRI scans, one important milestone was the release of a collected set of knee MRI scans and their annotations by Stanford scientists \cite{bien2018deep}. We utilized this data to conduct our comparative evaluations.  

The aim of this work is to further enhance the classification performance and thereby improve upon the clinical utility of using an artificially intelligent pipeline for predicting and diagnosing abnormalities in the knee region (using MR as the imaging modality). We use multi-view MRI scans and deep learning models to classify three commonly found abnormalities in knee (ACL and meniscus tears, and general abnormality). 
We have used multi-view magnetic resonance imaging knee scans for the classification of knee abnormality, ACL and meniscus tears with significant performance. Our experiments have shown that widely used deep learning architectures can be utilized under transfer learning to obtain highly accurate sensitivity and specificity without the need for very large scale medical imaging data.


\section{Methods}
The proposed methodology is illustrated in Figure \ref{fig}. We have used multiple deep learning models based on convolutional neural networks for classifying three different abnormalities from knee MRI. In our proposed method, we employed the transfer learning approach using pre-trained models on ImageNet \cite{dengimagenet}. 
In particular, we used the fine tuning strategy, where we updated all the parameters of the deep learning models for knee abnormality classification with weights initialized from the pre-trained models. 

\subsection{Pre-processing}
The input slices in three planes (sagittal, coronal, and axial) were extracted form DICOM images. All input images were normalized using the histogram based method for intensity standardization \cite{bien2018deep}. Each plane is considered as a 'view' in the classification network.  

\subsection{Deep learning models}
Deep learning models are contributing significantly to clinical radiology. In MSK radiology, in particular in knee related studies, it has been used in lesion detection (fractures, abnormalities in cartilage, ACL and meniscus tears), segmentation (bone and cartilage), classification (osteoarthritis grading, bone age assessment) and non-interpretive AI (reconstruction, resolution enhancement) \cite{chea2019current}. In our proposed method, we used deep learning models for classification of abnormalities in the knee region using multi-view MRI scans. For each architecture used in our study (AlextNet, ResNet-18, and GoogLeNet), a separate CNN model for each MRI plane (including sagittal, coronal, and axial) was trained. Since there are three classes, we trained a total of 9 CNN models for each architecture. The implementation details of these architectures are presented in the following. 

\subsection*{\textbf{AlexNet}} This architecture was proposed in 2012, and has been one of the significant milestones in the success of CNN based deep learning models. The original architecture was proposed using rectified linear unit (ReLU) as a non-linearity and dropout layers to achieve generalization \cite{krizhevsky2012imagenet}. The architecture consisted of five convolutional layers and three fully connected layers. The network was trained using the stochastic gradient descent algorithm.   
\subsection*{\textbf{ResNet}}
It has been shown that very deep networks have excellent performance in vision tasks, which was further improved by introducing residual connections \cite{he2016deep}. This has allowed using very deep networks with up to 152 layers having lesser complexity and better performance. Since ResNet has been traditionally very successful in learning patterns with in a data, we used ResNet-18 which is deeper and wider than AlexNet. 

\subsection*{\textbf{GoogLeNet}}
The architecture was proposed in 2014 and outperformed all other methods in the ILSVRC14 challenge \cite{szegedy2015going}. The architecture used inception modules and consisted of 22 layers. The architecture was aimed at having deeper and wider networks by adding minimal computational costs. We adopted this network to the knee abnormality classification task with weights initialized from the pre-trained model from ImageNet.      

\subsection{Abnormality Prediction/Classification}
For each of the classes such as abnormality, ACL and Meniscus, we trained 3 CNNs in totaling 9 CNNs on the training dataset. After training 3 CNNs for each abnormality, we trained three logistic regression models using 3 CNNs predictions on the training dataset. The results from multi-view MRIs were combined using logistic regression, which was trained using the training data to give a probability for each of the three abnormalities (shown in Figure~\ref{fig}). These three values gave us the probability/prediction of each of the abnormality (including ACL and meniscus tears as well as general abnormality). 

\section{Experiments and Results}

\subsection{Dataset}
We have used publicly available challenge dataset, named MRNet, comprising of multi-view knee MRIs \cite{bien2018deep}. The dataset consists of 1370 MRI exams from Stanford University medical center, MRI scans were acquired between 2001 - 2012. The imaging data includes 1,104 scans having abnormality, among which 508 had meniscal tears and 319 had ACL tears (194 scans had both meniscal and ACL tears). In our experiments, we used 1130 exams for training and 120 exams for testing. 

\begin{table*}[!t]
\centering
\caption{Summary of the classification performance using various deep learning models. In single-view model, only best results are reported for the sake of the space. Otherwise single-view models include three results corresponding to axial, coronal, and sagital. AUC: area under curve.}
\resizebox{!}{0.18\textwidth}{
 \begin{tabular}{|c|c|c|c|c|c|} 
 \hline
 \textbf{Deep Learning Model} & \textbf{Class}&\textbf{AUC} & \textbf{Sensitivity} & \textbf{Specificity}&\textbf{Accuracy}\\ [0.5ex] 
 \hline
 
 \hline
  \multirow{3}{*}{AlexNet-multiview}  & Abnormal & 0.8914 & 0.9789 & 0.4000 & 0.8583 \\ 
 & ACL & 0.9388 & 0.6852 & 0.9545 & 0.8333 \\
 & Meniscus & 0.8060 & 0.6923 & 0.8088 & 0.7583 \\ \cline{2-6}
 & \textbf{Average} & \textbf{0.8787} & \textbf{0.7855} & \textbf{0.7211} & \textbf{0.8166}\\
  \hline
   \multirow{1}{*}{AlexNet-single view}  & Average & 0.8387 & 0.6850 & 0.6939 & 0.7639 \\ \hline  
  \hline
  \multirow{3}{*}{ResNet-18-multiview}  & Abnormal & 0.8114 & 0.9684 &0.2800 & 0.8250\\ 
 & ACL & 0.9540 & 0.7778 &0.9394 & 0.8667\\
 & Meniscus & 0.8083 & 0.6346 &0.8529 & 0.7583 \\ \cline{2-6}
 & \textbf{Average} & \textbf{0.8579} & \textbf{0.7936} & \textbf{0.6908} & \textbf{0.8167}\\
  \hline
     \multirow{1}{*}{ResNet-18-single view}  & Average &0.7779  &0.7759  & 0.6136 & 0.7537 \\ \hline  
    \hline
  \multirow{3}{*}{GoogLeNet-multiview}  & Abnormal & 0.9091 & 0.9789 &0.2800 & 0.8333\\ 
 & ACL & 0.8906 & 0.6667 &0.9242 & 0.8083\\
 & Meniscus & 0.7791 & 0.6154 &0.7647 & 0.7000 \\ \cline{2-6}
 & \textbf{Average} & \textbf{0.8596} & \textbf{0.7537} & \textbf{0.6563} & \textbf{0.7806}\\
  \hline
       \multirow{1}{*}{GoogLeNet-single view}  & Average & 0.7605 & 0.6231 & 0.6847 & 0.7046 \\ \hline  
 \end{tabular}}
 \label{table:stats_all}
\end{table*}
\subsection{Network Parameters}

For all architectures that we comparatively evaluated, we avoided overfitting by using early stopping criteria in the training stage. A total of 50 epochs was sufficient for convergence in training. In the optimization stage, we used Adam optimizer with a weight decay of $0.1$ and a learning rate of 0.0001. We also used data augmentation as follows: each training example in the data was flipped horizontally with 0.5 \% probability, rotated randomly between $-25$ and $25$ degrees,  and shifted randomly between -25 and 25 pixels. 



\subsection{Performance Parameters}

We used area under the curve (AUC), accuracy, sensitivity, and specificity to evaluate our experiments. The classification performance of the models 
is presented in Table \ref{table:stats_all}. 
All experiments were conducted on  an NVIDIA Titan Xp graphical processing unit with 12GB memory

 


\subsection{Results and Discussion}
The summary of results (Table \ref{table:stats_all}) for various deep learning models indicate that a significant performance was achieved for all three classes. On an average (for all three classes), the AUC was 0.8787, 0.8579, and 0.8596 for AlexNet, ResNet-18, and GoogLeNet, respectively. The highest accuracy for abnormality was 85.83\% using AlexNet, 86.67\% for ACL using ResNet-18, and 75.83\% using both AlexNet and ResNet-18 architectures. In general, ResNet-18 out-performed AlexNet and GoogLeNet in all performance parameters except in specificity (0.6908). We also present the results when single view MRIs are used. It is observed that the performance when using multiview MRIs outperforms the models trained on single view MRIs.

The overall results showed that meniscus tear was the most challenging class with the lower performance in terms of sensitivity. One way to improve the classification performance with more advanced architectures such as DenseNet and CapsNet. However, exploring incrementally more advanced architectures is beyond the scope of this paper. One alternative way to obtain improved classification results maybe to create a deep learning model relying on more imaging data with clear labels and training the whole system from scratch. In medical imaging based deep learning applications, however, obtaining explicitly labeled (clean) large amount of imaging data is an extreme challenge. 

\section{Conclusion}
The need for an automated method for analyzing knee MRIs has recently been augmented with the number of knee scans performed in MSK radiology. At the same time, improvement in scanning hardware has led to generation of radiology images with a lot of clinical details. This would benefit in clinical prognosis and diagnosis but has increased the burden on radiologists with increasing number of scans to analyze. The development of automated methods that can supplement radiologists in interpretation and decision making using radiology scans is the way forward. Towards this overarching goal, we have used deep learning models for the analysis of knee MRIs for classifying abnormalities, meniscal tears and ACL tears. Our results indicate that deep learning image analysis methods could solve challenging problems in MSK radiology.         
\section{Acknowledgment}

We would like to thank NVIDIA for donating a Titan Xp GPU for our experiments.


\bibliographystyle{IEEEtran}
\bibliography{IEEEexample}


\end{document}